\def\monthyear{\ifcase\month\or
  January\or February\or March\or April\or May\or June\or
  July\or August\or September\or October\or November\or
December\fi
  \space\number\year}

\renewcommand{\baselinestretch}{1.17}
\def\section{\@startsection {section}{1}{\z@}{-3.5ex plus -1ex
minus
 -.2ex}{2.3ex plus .2ex}{\large\bf}}
\def\subsection{\@startsection{subsection}{2}{\z@}{-3.25ex plus -
1ex minus
 -.2ex}{1.5ex plus .2ex}{\normalsize\bf}}

\newcommand{\gapproxeq}{\lower
.7ex\hbox{$\;\stackrel{\textstyle >}{\sim}\;$}}
\newcommand{\lapproxeq}{\lower
.7ex\hbox{$\;\stackrel{\textstyle <}{\sim}\;$}}

\newcounter{appendice}

\def\thefiglist#1{\section*{Figure Captions\markboth
 {FIGURE CAPTIONS}{FIGURE CAPTIONS}}\list
 {Figure \arabic{enumi}.}
 {\settowidth\labelwidth{Figure #1.}\leftmargin\labelwidth
 \advance\leftmargin\labelsep
 \usecounter{enumi}}
 \def\baselinestretch{1.1}\@normalsize
 \def\newblock{\hskip .11em plus .33em minus -.07em}
 \sloppy}

\documentstyle[12pt]{article}

\newcommand{\Tr}{\,\mbox{\rm Tr}}

\newcommand{\hodge}{\mbox{\mbox{}$^*\!$}}
\begin{document}
\begin{titlepage}
\begin{flushright} RAL-93-080\\
hep-th/9312072
\end{flushright}
\vskip 1cm
\begin{center}
{\bf\Large Tensor Gauge Potentials in Loop Space Formulation of Yang-Mills
Fields}
\vskip 1cm
{\large Chan Hong-Mo}\\
\vskip .3cm
{\it Rutherford Appleton Laboratory\\
Chilton, Didcot, Oxon, OX11 0QX, UK.}
\vskip .5cm
{\large J. Faridani}\\
\vskip .3cm
{\it Department of Theoretical Physics, Oxford University,\\
1 Keble Road, Oxford, OX1 3NP, UK.}
\vskip .5cm
{\large Tsou Sheung Tsun}\\
\vskip .3cm
{\it Mathematical Institute, Oxford University,\\
24-29 St. Giles, Oxford, OX1 3LB, UK.}\\
\end{center}
\begin{abstract}
It is shown that an antisymmetric rank-two tensor gauge potential
of the type first found in string and supersymmetry theories occurs also
in ordinary Yang-Mills theory when formulated in loop space, where it
appears as a Lagrange multiplier for a zero curvature constraint necessary
and sufficient for removing the inherent redundancy of loop variables.  It
is then further shown that the tensor potential acts there as the parallel
`phase' transport for monopoles.
\end{abstract}
\end{titlepage}

Recently there has been renewed interest in antisymmetric tensor gauge
potentials \cite{Dealwisetal,Batllegomis,Clarketal,Blagojevic,Thierrymieg,
Blauthompson,Ariasetal}, especially in the `nonabelian'
version suggested by Freedman and Townsend in 1981 \cite{Freetown}.
Although the present emphasis is on the quantization
of the theory, the physical or geometrical significance of these potentials
remain somewhat obscure in that it is unclear whether they function as
connections for parallel `phase' transport as ordinary vector gauge
potentials do, and if they do so, what the `phase' is that they transport.
In this paper, we wish to point out first that, though originally discovered
in string and supersymmetric theories, these tensor potentials occur also
in ordinary Yang-Mills theory when formulated in loop space, where they
appear as Lagrange multipliers for the constraint needed to remove the
intrinsic redundancy of loop variables, and second, that in this context
they function as the parallel transport of the `phases' of monopoles.  It
is hoped that these observations will help with the general understanding
of the physical significance of tensor gauge potentials.

Yang-Mills theory can be fully described by means of the loop variables:
\begin{equation}
F_\mu[\xi|s] = \frac{i}{g} \Phi[\xi]^{-1} \frac{\delta}{\delta \xi^\mu(s)}
   \Phi[\xi],
\label{Fmuxis}
\end{equation}
introduced by Polyakov \cite{Polyakov1}, where $\Phi[\xi]$ is the phase factor:
\begin{equation}
\Phi[\xi] = P_s \exp ig \int_0^{2\pi} A_\mu(\xi(s)) \dot{\xi}^\mu(s) ds.
\label{Phiofxi}
\end{equation}
In (\ref{Phiofxi}), a dot denotes differentiation with respect to the loop
parameter $s$, $P_s$ denotes ordering in $s$, from right to left for
increasing $s$ in our convention, and $\xi$ represents a parametrized loop
passing through a fixed reference point $P_0$, namely:
\begin{equation}
\xi = \{\xi(s), s = 0 \rightarrow 2\pi, \xi(0) = \xi(2\pi) = P_0 \}.
\label{xiofs}
\end{equation}

In terms of ordinary space-time variables, $F_\mu[\xi|s]$ can be expressed
as:
\begin{equation}
F_\mu[\xi|s] = \Phi_\xi^{-1}(s,0) F_{\mu\nu}(\xi(s)) \dot{\xi}^\nu(s)
    \Phi_\xi(s,0),
\label{Fmuxisinx}
\end{equation}
where:
\begin{equation}
\Phi_\xi(s_2,s_1) = P_s \exp ig \int_{s_1}^{s_2} A_\mu(\xi(s))
   \dot{\xi}^\mu(s) ds
\label{Phixis1s2}
\end{equation}
is the parallel transport from $\xi(s_1)$ to $\xi(s_2)$ along the loop $\xi$.

{}From (\ref{Fmuxisinx}), it follows that $F_\mu[\xi|s]$ can depend on the loop
coordinate $\xi(s')$ only for $s' \leq s$, i.e.:
\begin{equation}
\frac{\delta}{\delta \xi^\nu(s')} F_\mu[\xi|s] = 0, \ \ \ s'>s,
\label{xiuptos}
\end{equation}
which is a consequence of our ordering convention $P_s$ in (\ref{Phiofxi}),
and that it has only components transverse to the loop, namely that:
\begin{equation}
F_\mu[\xi|s] \dot{\xi}^\mu(s) = 0,
\label{transverseF}
\end{equation}
which is equivalent to the fact that $\Phi[\xi]$ is by definition independent
of how the loop $\xi$ is parametrized.  These two properties (\ref{xiuptos})
and (\ref{transverseF}) which in effect just reduce the range of the arguments
and the number of components of $F_\mu[\xi|s]$ will henceforth be regarded
as understood and absorbed into the notation.

The variables $F_\mu[\xi|s]$, like the phase factors $\Phi[\xi]$, are gauge
invariant except for an $x$-independent rotation at the reference point $P_0$
which is easily handled.  But, like $\Phi[\xi]$ also, they form a highly
redundant set and have to be severely constrained.  By the latter assertion,
we mean that in order effectively to employ $F_\mu[\xi|s]$ as variables to
describe Yang-Mills theory, which we know is already adequately described by
$A_\mu(x)$, we have to ensure that the loop variables can indeed be expressed
in terms of some vector potential $A_\mu(x)$ in the manner of (\ref{Fmuxis})
and (\ref{Phiofxi}).  However, this may not be true for just any given set of
$F_\mu[\xi|s]$.

The conditions that $F_\mu[\xi|s]$ have to satisfy for (\ref{Fmuxis}) and
(\ref{Phiofxi}) to hold is best stated in terms of the quantity
\cite{Polyakov1}:
\begin{equation}
G_{\mu\nu}[\xi|s] = \frac{\delta}{\delta \xi^\nu(s)} F_\mu[\xi|s] -
   \frac{\delta}{\delta \xi^\mu(s)} F_\nu[\xi|s]+ig[F_\mu[\xi|s],F_\nu[\xi|s]].
\label{Gmunuxis}
\end{equation}
{}From its definition (\ref{Fmuxis}), we note that $F_\mu[\xi|s]$ can be
interpreted as a connection in loop space for parallel transport of the
phase $\Phi[\xi]$.  In this sense then, $G_{\mu\nu}[\xi|s]$ is the
corresponding curvature.  Now it can readily be seen that so long as
$F_\mu[\xi|s]$ is expressible in terms of an $A_\mu(x)$ through (\ref{Phiofxi})
and (\ref{Fmuxis}), then it satisfies the following condition:
\begin{equation}
G_{\mu\nu}[\xi|s] = 0,
\label{Gausslaw0}
\end{equation}
meaning that, as a connection, it is pure gauge, giving thus zero curvature.
Conversely, it can also be shown, though less readily and perhaps for this
reason less widely recognized, that provided $F_\mu[\xi|s]$ satisfies
(\ref{Gausslaw0}), (\ref{xiuptos}) and (\ref{transverseF}) being understood,
then there will exist an $A_\mu(x)$ in terms of which $F_\mu[\xi|s]$ can be
expressed through (\ref{Fmuxis}) and (\ref{Phiofxi}) as required
\cite{Chanstsou1}.  In other words, the constraint (\ref{Gausslaw0}) removes
exactly the redundancy of the loop variables $F_\mu[\xi|s]$ and makes a
description of the theory in terms of them equivalent to the original
description in terms of the gauge potential $A_\mu(x)$.

{}From (\ref{Fmuxisinx}), it follows that the Yang-Mills action:
\begin{equation}
{\cal A}^0_F = -\frac{1}{16\pi} \int \!d^4x \Tr\{F_{\mu\nu}(x) F^{\mu\nu}(x)\}
\label{freeactF}
\end{equation}
can be rewritten in terms of loop variables as:
\begin{equation}
{\cal A}^0_F = -\frac{1}{4\pi {\bar N}} \int \delta \xi \int_0^{2\pi} \!ds
   \Tr\{F_\mu[\xi|s] F^\mu[\xi|s]\} \dot{\xi}^{-2}(s),
\label{freeactFl}
\end{equation}
where ${\bar N}$ is a normalization factor:
\begin{equation}
{\bar N} = \int_0^{2\pi} ds \int \prod_{s' \neq s} d^4 \xi(s').
\label{Nbar}
\end{equation}
The variables $F_\mu[\xi|s]$, however, are not independent variables, but,
as stated above, have to satisfy the constraint (\ref{Gausslaw0}) for all
$\xi$ and $s$.  Hence, introducing the Lagrange multipliers $L_{\mu\nu}[\xi|s]$
antisymmetric in $\mu,\nu$, and incorporating the constraint into the action,
we have:
\begin{equation}
{\cal A}_F = {\cal A}^0_F + \int \delta \xi ds \Tr \{L^{\mu\nu}[\xi|s]
   G_{\mu\nu}[\xi|s] \}.
\label{freeactFlc}
\end{equation}
For example, to find the field equations, we extremize (\ref{freeactFlc})
with respect to $F_\mu[\xi|s]$, obtaining:
\begin{equation}
(4\pi {\bar N} \dot{\xi}^2(s))^{-1} F_\mu[\xi|s]
   = -{\cal D}^\nu(s) L_{\mu\nu}[\xi|s],
\label{eleqfreeF}
\end{equation}
where
\begin{equation}
{\cal D}_\nu(s) = \frac{\delta}{\delta \xi^\nu(s)} - ig [F_\nu[\xi|s], \ \ \ ]
\label{curlyD}
\end{equation}
is a kind of covariant derivative in loop space with $F_\mu[\xi|s]$ as
connection.  From (\ref{eleqfreeF}), one deduces that:
\begin{equation}
(2\pi {\bar N} \dot{\xi}^2(s))^{-1} \frac{\delta}{\delta \xi_\mu(s)}
   F_\mu[\xi|s] = - [{\cal D}^\mu(s), {\cal D}^\nu(s)] L_{\mu\nu}[\xi|s],
\label{propolyakov}
\end{equation}
where the right-hand side is just $ig[G^{\mu\nu}[\xi|s], L_{\mu\nu}[\xi|s]]$,
and hence from the constraint (\ref{Gausslaw0}) one obtains:
\begin{equation}
\frac{\delta}{\delta \xi_\mu(s)} F_\mu[\xi|s] = 0,
\label{Polyakov}
\end{equation}
which, as pointed out by Polyakov \cite{Polyakov1}, is the loop space
statement of the Yang-Mills equation.

One sees that apart from some trivial changes in notation, (\ref{freeactFlc})
is exactly the loop space version of the first order Freedman-Townsend action
\cite{Freetown} for a tensor gauge field with $L_{\mu\nu}[\xi|s]$ as the
potential.  Indeed, under the transformation:
\begin{equation}
\Delta F_\mu[\xi|s] = 0,
\label{DeltaF}
\end{equation}
and:
\begin{equation}
\Delta L_{\mu\nu}[\xi|s] = \epsilon_{\mu\nu\rho\sigma}
   {\cal D}^\rho(s)\Lambda^\sigma[\xi|s],
\label{DeltaL}
\end{equation}
the action (\ref{freeactFlc}) is invariant  for arbitrary functions
$\Lambda_\mu[\xi|s]$ by virtue of the Bianchi identity of $G_{\mu\nu}[\xi|s]$.
We notice then that the dual of $L_{\mu\nu}[\xi|s]$ defined by:
\begin{equation}
\hodge L_{\mu\nu}[\xi|s] = -(1/2)
\epsilon_{\mu\nu\rho\sigma} L^{\rho\sigma}[\xi|s]
\label{Lmunustar}
\end{equation}
will transform as:
\begin{equation}
\Delta \hodge L_{\mu\nu}[\xi|s] = {\cal D}_\nu(s) \Lambda_\mu[\xi|s]
   - {\cal D}_\mu(s) \Lambda_\nu[\xi|s],
\label{DeltaLstar}
\end{equation}
exactly as the tensor gauge potential in Freedman and Townsend.

The question now is: what does this transformation represent?  It is not the
original Yang-Mills gauge transformation which has already been removed by
going over into the gauge-invariant (apart from a trivial $x$-independent
transformation at the reference point $P_0$) loop variables $F_\mu[\xi|s]$.
What we shall show in fact is that it represents a local change in `phase' of
`colour' monopoles in the Yang-Mills fields, in much the same way that the
original Yang-Mills gauge transformation represents a local change in `phase'
of `colour' sources.  Further, it is for this `phase' of `colour' monopoles
that
the tensor potential $L_{\mu\nu}[\xi|s]$ is acting as parallel transport.

We should first make clear that by `colour' monopoles of Yang-Mills fields,
we do not mean here the 't Hooft-Polyakov soliton solutions of 1974
\cite{tHooft,Polyakov2} which are abelian monopoles embedded in a nonabelian
Yang-Mills-Higgs field.  We mean rather the generalization of the Dirac point
monopole \cite{Dirac} to nonabelian theories as first suggested by Lubkin
\cite{Lubkin}, Wu-Yang \cite{Wuyang} and Coleman \cite{Coleman}.  These latter
are characterized by nontrivial nonabelian bundles over $S^2$,
whereas the former are characterized by nontrivial
abelian subbundles of a trivial nonabelian bundle as in the original
't Hooft-Polyakov papers.  In this language then, the charge of a `colour'
monopole in a Yang-Mills theory with gauge group $G$ takes values in the
fundamental group $\pi_1(G)$.  In particular, for the simplest pure Yang-Mills
theory with gauge algebra ${\bf su}(2)$ and gauge group $G = SO(3)$, the
monopole charge $\zeta$ can take only values in ${\bf Z}_2$ and can thus be
labelled just by a sign $\pm$ \cite{Chantsou}.

The monopole charge so defined which is enclosed inside any given surface
$\Sigma$ is given by the holonomy over this surface \cite{Chanstsou1}:
\begin{equation}
\Theta_\Sigma = \zeta_\Sigma,
\label{Gausslawli}
\end{equation}
where the surface $\Sigma$ passing through the loop-space reference point
$P_0 = \{\xi_0^\mu\}$ is considered as a closed loop in loop space.  This
holonomy $\Theta_\Sigma$ can be written explicitly as:
\begin{equation}
\Theta_\Sigma = P_t \exp ig \int_0^{2\pi} dt \int_0^{2\pi} ds F_\mu[\xi_t|s]
   \frac{\partial\xi_t^\mu(s)}{\partial t}
\label{ThetainF}
\end{equation}
for any parametrization $\{\xi_t^\mu(s)\}$ of $\Sigma$:
\begin{equation}
\Sigma = \{\xi_t(s); s = 0 \rightarrow 2\pi, t = 0 \rightarrow 2\pi,
   \xi_t(0) = \xi_t(2\pi) = \xi_0(s) = \xi_{2\pi}(s) = \xi_0 \}.
\label{Sigmapar}
\end{equation}
For example, in the pure ${\bf su}(2)$ Yang-Mills theory with gauge group
$SO(3)$, the quantity in (\ref{ThetainF}) for any $\Sigma$ will be
an element of the group $SU(2)$ taking the values $\pm I$ in its centre
${\bf Z}_2 = SU(2)/SO(3)$.  The value $-I$ for $\Theta_\Sigma$ will then
signify that there is a `colour' monopole enclosed inside the surface
$\Sigma$.

The existence of a monopole charge at some space-time point $x$ means that
the loop space curvature $G_{\mu\nu}[\xi|s]$ of (\ref{Gmunuxis}) will fail to
vanish at $\xi(s) = x$. In other words, a monopole charge may be interpreted as
a source of loop space curvature \cite{Chanstsou1,Chanstsou2}.  This is
not in contradiction with the statement above in (\ref{Gausslaw0}) since
the presence of a monopole necessitates patching of the gauge potential
so that at the position of the monopole $A_\mu(x)$ is not defined, violating
thus the conditions for (\ref{Gausslaw0}) to hold.  Indeed, since geometrically
the curvature is just a differential version of the holonomy, it follows
from (\ref{Gausslawli}) that at the monopole position, $G_{\mu\nu}[\xi|s]$
must take some value $4\pi{\tilde g}\kappa$ in the gauge Lie algebra where
$\kappa$ satisfies:
\begin{equation}
\exp i\pi \kappa = \zeta
\label{defkappa}
\end{equation}
for a monopole of charge $\zeta$.  Thus, in particular, for a monople of
charge $-$ in the $SO(3)$ theory, $\kappa$ will take a value $n\sigma$
for $n$ odd and $\sigma = \alpha_i \tau^i$, where $\alpha$ is a unit vector and
$\tau^i$ are the Pauli matrices.  Notice that in ascribing the algebra element
$\kappa$ to the monopole, one has assigned to it a `phase' or orientation
in internal symmetry space which it did not originally possess.

Suppose now that there is a monopole moving along a world-line $Y(\tau)$,
then (\ref{Gausslaw0}) will be replaced by:
\begin{equation}
G_{\mu\nu}[\xi|s] = 4\pi {\tilde g} \kappa[\xi|s] \epsilon_{\mu\nu\rho\sigma}
   \dot{\xi}^\rho(s) \int d\tau \frac{dY^\sigma(\tau)}{d\tau}
   \delta^4(Y(\tau)-\xi(s)).
\label{Gausslawld}
\end{equation}
It can be shown that provided this is satisfied, the existence of the gauge
potential $A_\mu(x)$ is still guaranteed at all points in space-time
\cite{Chanstsou1}, except of course on the monopole world-line where we
cannot expect the potential to be defined in any case.  Hence, the condition
(\ref{Gausslawld}) will again remove the redundancy from the loop variables
as (\ref{Gausslaw0}) did before, meaning now that for all loops not passing
through $Y(\tau)$, it is assured that we may still write (\ref{Phiofxi})
and (\ref{Fmuxis}) as before.

Replacing (\ref{Gausslaw0}) by (\ref{Gausslawld}) as constraint in
(\ref{freeactFlc}), however, means that there is coupling now between the
field as represented by $F_\mu[\xi|s]$ and the monopole coordinate $Y(\tau)$
or that there is an induced interaction between the field and the monopole.
Indeed, this seems a natural way to define the interaction and it has been
shown that in this way equations of motion for the field-monopole system can be
derived which for the abelian theory reduce exactly to the Maxwell and Lorentz
equations for the magnetic charge \cite{Chanstsou2,Chanftsou}.

Our primary interest here, however, is not monopole dynamics but the meaning
of the antisymmetric tensor potential $L_{\mu\nu}[\xi|s]$ as connection.  To
see this, we shall need to extend the discussion to a quantum monopole
described by a wave function so that we may follow the variation of the
wave function's `phase' under parallel transport as we did for the electron
wave function in the Bohm-Aharonov experiment \cite{Bohmanov}.

We notice that the right-hand side of (\ref{Gausslawld}) is basically just
the monopole current.  Indeed, in the special case of an abelian theory,
the equation reduces by virtue of (\ref{Fmuxisinx}) simply to:
\begin{equation}
\partial_\nu \hodge F^{\mu\nu}(x) = -4\pi {\tilde e} \int d\tau
   \frac{dY^\mu(\tau)}{d\tau} \delta^4(x-Y(\tau)),
\label{gausslawd}
\end{equation}
in which, with ${\tilde e}$ being the magnetic charge, it is exactly the
magnetic current which appears on the right-hand side.  In replacing a
classical monopole by a quantum monopole described by a (say Dirac) wave
function, the classical current on the right-hand side of (\ref{gausslawd})
is replaced by the quantum current, so that instead of (\ref{gausslawd})
we have:
\begin{equation}
\partial_\nu \hodge F^{\mu\nu}(x) = -4\pi {\tilde e}
{\bar \psi}(x) \gamma^\mu   \psi(x).
\label{gausslawdq}
\end{equation}
Equivalently, in loop space notation, we have:
\begin{equation}
G_{\mu\nu}[\xi|s] = 4\pi {\tilde e} \epsilon_{\mu\nu\rho\sigma}
   \dot{\xi}^\rho(s) {\bar \psi}(\xi(s)) \gamma^\sigma \psi(\xi(s)).
\label{gausslawldq}
\end{equation}
For the general Yang-Mills case, the corresponding constraint reads as
\cite{Chanftsou}:
\begin{equation}
G_{\mu\nu}[\xi|s] = 4\pi {\tilde g} \epsilon_{\mu\nu\rho\sigma}
   \dot{\xi}^\rho(s) \Omega^{-1}_\xi(s,0) \{ [{\bar \psi}(\xi(s))
   \gamma^\sigma \tau^i \psi(\xi(s)] \tau_i \} \Omega_\xi(s,0),
\label{Gausslawldq}
\end{equation}
where the `current' inside the curly brackets is again an element of the
gauge Lie algebra with a `phase' or orientation in internal symmetry space.
This `phase', however, is measured in the local monopole frame at $\xi(s)$,
hence the rotation matrices $\Omega_\xi(s,0)$ to transform back to the
reference frame at $\xi_0$ in which $G_{\mu\nu}[\xi|s]$ on the left-hand
side is measured.

Suppose now we write an action for the field-monopole system.  We shall
then write the free action as usual as:
\begin{equation}
{\cal A}^0 = {\cal A}^0_F + \int d^4x \,
{\bar \psi}(x) (i\partial_\mu \gamma^\mu - m) \psi(x),
\label{freeact}
\end{equation}
where we may again express ${\cal A}^0_F$ in terms of the loop variables
$F_\mu[\xi|s]$ as in (\ref{freeactFl}).  The loop variables have again to
be constrained, but now by (\ref{Gausslawldq}) instead of (\ref{Gausslaw0}).
Incorporating then the constraint into the action by means of Lagrange
multipliers $L_{\mu\nu}[\xi|s]$ as before, we have:
\begin{equation}
{\cal A} = {\cal A}^0 + \int \delta \xi ds \Tr \{L^{\mu\nu}[\xi|s]
   (G_{\mu\nu}[\xi|s] + 4\pi J_{\mu\nu}[\xi|s]) \},
\label{fullact}
\end{equation}
where $J_{\mu\nu}[\xi|s]$ denotes $-1/4\pi$ times the right-hand side of
(\ref{Gausslawldq}).

We ask now what happens if we perform the transformations (\ref{DeltaF}) and
(\ref{DeltaL}) on ${\cal A}$.  We have seen already that the field part of
the action remains invariant.  Next, let us examine the last term on the
right of (\ref{fullact}), namely:
\begin{equation}
4 \pi \int \delta \xi ds \Tr \{ L^{\mu\nu}[\xi|s] J_{\mu\nu}[\xi|s] \}
   = {\tilde g} \int d^4x {\bar \psi}(x) {\tilde A}_\mu(x) \gamma^\mu \psi(x),
\label{tinteract}
\end{equation}
where, from the definition of $J_{\mu\nu}[\xi|s]$ as the right-hand side
of (\ref{Gausslawldq}):
\begin{equation}
{\tilde A}_\mu(x) = - 8\pi \int \delta \xi ds
   \Omega_\xi(s,0) \hodge L_{\mu\nu}[\xi|s] \Omega^{-1}_\xi(s,0)
   \dot{\xi}^\nu(s) \delta^4(x-\xi(s)).
\label{Atildemu}
\end{equation}
The transformation (\ref{DeltaL}) will give an increment to ${\tilde A}_\mu(x)$
of the form:
\begin{equation}
\Delta_L{\tilde A}_\mu(x) = - 8\pi \int \delta \xi ds \Omega_\xi(s,0)
   {\cal D}_\mu(s) \Lambda_\nu[\xi|s] \Omega^{-1}_\xi(s,0)
   \dot{\xi}^\nu(s) \delta^4(x-\xi(s)),
\label{DeltaLAtil}
\end{equation}
where we have used the fact that loop quantities have no derivatives
longitudinal to the loop because of reparametrization invariance.  Next, using
the Bianchi identity satisfied by $G_{\mu\nu}[\xi|s]$, we can deduce from
(\ref{Gausslawld}) and the conservation of the monopole current that
\cite{Chanftsou}:
\begin{equation}
\frac{\delta}{\delta \xi^\mu(s)} \Omega_\xi(s,0)
   = -ig \Omega_\xi[\xi|s] F_\mu[\xi|s],
\label{difOmega}
\end{equation}
so that:
\begin{equation}
\frac{\delta}{\delta \xi^\mu(s)} \{\Omega_\xi(s,0) \Lambda_\nu[\xi|s]
   \Omega^{-1}_\xi(s,0) \} = \Omega_\xi(s,0) \{ {\cal D}_\mu(s)
   \Lambda_\nu[\xi|s] \} \Omega^{-1}_\xi(s,0).
\label{cordifOmega}
\end{equation}
Substituting this into (\ref{DeltaLAtil}), one obtains after some manipulation:
\begin{equation}
\Delta_L{\tilde A}_\mu(x) = \partial_\mu {\tilde \Lambda}(x),
\label{DeltaLAtila}
\end{equation}
where:
\begin{equation}
{\tilde \Lambda}(x) = - 8 \pi \int \delta \xi ds \Omega_\xi(s,0)
   \Lambda_\nu[\xi|s] \Omega^{-1}_\xi(s,0) \dot{\xi}^\nu(s)
   \delta^4(x-\xi(s)).
\label{Lambdatil}
\end{equation}

Such an increment in ${\tilde A}_\mu(x)$ will induce a corresponding change
in the action ${\cal A}$ of (\ref{fullact}), which will however be cancelled
if we perform at the same time the following transformation on the monopole
wave function $\psi(x)$:
\begin{equation}
\psi(x) \longrightarrow (1 + i{\tilde g} {\tilde \Lambda}(x)) \psi(x),
\label{psitransf}
\end{equation}
meaning a rotation of the local `phase' of the monopole, since this
will induce a similar rotation in the frame-transformation factors
$\Omega_\xi(s,0)$, thus:
\begin{equation}
\Omega_\xi(s,0) \longrightarrow [1 + i{\tilde g} {\tilde \Lambda}(\xi(s))]
   \Omega_\xi(s,0),
\label{Omegatransf}
\end{equation}
which gives the total transformation of ${\tilde A}_\mu(x)$ the standard
transformation for a nonabelian potential:
\begin{equation}
\Delta {\tilde A}_\mu(x) = \partial_\mu {\tilde \Lambda}(x)
   +i{\tilde g} [{\tilde \Lambda}(x), {\tilde A}_\mu(x)].
\label{DeltaAtil}
\end{equation}
The increment from the term (\ref{tinteract}) in the action (\ref{fullact})
will therefore cancel in a familiar manner with the variation in the free
action term (\ref{freeact}) from the transformation (\ref{psitransf}) of
the wave function $\psi(x)$.

One sees thus that the transformation (\ref{DeltaL}) of the tensor potential
$L_{\mu\nu}[\xi|s]$ does indeed correspond to a `phase' rotation of the
monopole wave function, and that the action for the field-monopole system
is invariant under this transformation.  We stress again that this is not the
original Yang-Mills gauge invariance which has already been absorbed into
the formulation by adopting the `gauge-invariant' loop quantities
$F_\mu[\xi|s]$
as variables.  The total gauge symmetry is now therefore $SU(N) \times SU(N)$
where the second $SU(N)$ carries (because of the $\epsilon$ symbol occuring
in, for example, (\ref{Gausslawldq})) a parity opposite to that of the first,
original Yang-Mills $SU(N)$.

What is the origin of this new additional gauge symmetry?  We recall that the
monopole charge was originally defined as an element in the centre of the
group $SU(N)$; in particular for $SU(2)$, it is labelled only by a sign $-$.
It does not therefore have at first a `phase'.  But when its dynamics
was formulated through the imposition of the constraint (\ref{Gausslawld}) or
(\ref{Gausslawldq}) as was done above, it was obliged to make a choice of
`phase', since the constraints imposed are equations in the algebra.
This choice of `phase' is `local', depending not only on the loop $\xi$, but
also on the `end-point' labelled by the parameter $s$.  However, the actual
physics cannot depend on this choice of `phase' since the monopole
charge itself has none.  Hence the dynamics must be invariant under an
arbitrary rotation in `phase' at every $\xi$ and $s$, which is indeed the
freedom enjoyed by the gauge parameter $\Lambda_\mu[\xi|s]$ in the symmetry
transformation (\ref{DeltaL}) found above.

Since one is allowed to rotate the monopole `phase' arbitrarily for each
$\xi$ and $s$, one needs a parallel transport or `connection' to specify
what is meant by the same `phase' at different values of $\xi$ and $s$,
namely to play the role here of the gauge potential $A_\mu(x)$ in the
original Yang-Mills symmetry.  This is provided by the tensor potential
$L_{\mu\nu}[\xi|s]$, which depends on $\xi$ and $s$ as expected.  The reason
why it should carry two indices $\mu,\nu$ instead of just one index $\mu$
as in the ordinary gauge potential $A_\mu(x)$ is that the monopole charge
as defined above is specified by a closed 2-dimensional surface enclosing
a 3-volume, and a 3-volume element in 4-dimensional space has a direction.
For this reason, the measured monopole charge depends not only on the
position but also on a direction in space-time.  The parallel transport
has therefore to specify `phase' variations not only for neighbouring
positions but also for `neighbouring' directions, hence the extra index.

Although our observations here on the tensor potential are made specifically
only within the framework of the loop space formulation of Yang-Mills fields,
it is hoped that they will be useful also for understanding the geometrical
meaning of tensor potentials in general.

\subsection*{Acknowledgement}
We are indebted to Graham J. Ward for bringing to our notice Freedman and
Townsend's paper and the similarity of its tensor potential to ours.
One of us (JF) acknowledges the support of the Mihran and Azniv Essefian
Foundation (London), the Soudavar Foundation (Oxford) and the Calouste
Gulbenkian Foundation (Lisbon), while another (TST) thanks the Windgate
Foundation for partial support during the latter part of this work.


\begin{thebibliography}{99}
\bibitem{Dealwisetal} S.P. De Alwis, M.T. Grisaru and L. Mezincescu, Nucl.
   Phys. {\bf B303} (1988) 57.
\bibitem{Batllegomis} C. Batlle and J. Gomis, Phys. Rev. {\bf D38} (1988) 1169.
\bibitem{Clarketal} T.E. Clarke, C-H Lee and S.T. Love, Nucl. Phys. {\bf B308}
   (1988) 379.
\bibitem{Blagojevic} M. Blagojevic, D.S. Popovic and B. Sazdovic, Phys. Lett.
   {\bf 214B} (1988) 47.
\bibitem{Thierrymieg} J. Thierry-Mieg, Nucl. Phys. {\bf B335} (1990) 334.
\bibitem{Blauthompson} M. Blau and G. Thompson, Ann. Phys. {\bf 205} (1991)
137.
\bibitem{Ariasetal} P.J. Arias et al., Int. J. Mod. Phys. {\bf A4} (1992) 737.
\bibitem{Freetown} D.Z. Freedman and P.K. Townsend, Nucl. Phys. {\bf B177}
   (1981) 282.
\bibitem{Polyakov1} A.M. Polyakov, Nucl. Phys. {\bf 164} (1979) 171.
\bibitem{Chanstsou1} Chan Hong-Mo, P. Scharbach, and Tsou Sheung Tsun,
   Ann. Phys. (N.Y.) {\bf 166} (1986) 396.
\bibitem{tHooft} G. t' Hooft, Nucl. Phys. {\bf B79} (1974) 276.
\bibitem{Polyakov2} A.M. Polyakov, Sov. Phys. JETP {\bf 41} (1975) 988.
\bibitem{Dirac} P.A.M. Dirac, Proc. Roy. Soc. London {\bf A133} (1931) 60.
\bibitem{Lubkin} E. Lubkin, Ann. Phys. (N.Y.) {\bf 23} 233.
\bibitem{Wuyang} Tai Tsun Wu and Chen Ning Yang, Phys. Rev. {\bf D12} (1975)
   3845.
\bibitem{Coleman} S. Coleman, Erice School (1974) 297.
\bibitem{Chantsou} For an introduction to nonabelian monopoles, see e.g.
   Chan Hong-Mo and Tsou Sheung Tsun,
   {\it Some Elementary Gauge Theory Concepts}
   (World Scientific, 1993).
\bibitem{Chanstsou2} Chan Hong-Mo, P. Scharbach and Tsou Sheung Tsun,
   Ann. Phys. (N.Y.) {\bf 167} 454.
\bibitem{Chanftsou} Chan Hong-Mo, Tsou Sheung Tsun and J. Faridani, Rutherford
   Appleton Laboratory preprint (1993), RAL-93-079.
\bibitem{Bohmanov} Y. Aharonov and D. Bohm, Phys. Rev.{\bf 115} (1959) 485.
\end{thebibliography}
\end{document}